\documentclass[floatfix,prl,superscriptaddress,twocolumn,english,showpacs,letterpaper]{revtex4}
\usepackage{graphicx}
\usepackage{amsmath}
\usepackage{amssymb}
\usepackage{color}
\usepackage{wasysym}
\makeatletter

\begin{document}

\title{Chaos and universality in two-dimensional Ising
spin glasses}
\author{Creighton K. Thomas}
\affiliation{Department of Physics and Astronomy, Texas A\&M University,
College Station, Texas 77843-4242, USA}
\author{David A. Huse}
\affiliation{Department of Physics, Princeton University, Princeton, NJ 08544, USA}
\affiliation{Institute for Advanced Study, Princeton, NJ 08540, USA}
\author{A. Alan Middleton}
\affiliation{Department of Physics, Syracuse University, Syracuse, NY 13244, USA}
\begin{abstract}
Recently extended precise numerical
methods and droplet scaling arguments allow for a coherent
picture of the glassy states of two-dimensional Ising spin glasses to be assembled.
The length scale at which entropy becomes important and produces
``chaos'', the extreme sensitivity of the state to temperature, is found to depend on the type of randomness.
For the $\pm J$ model this length scale dominates the low-temperature specific heat.
Although there is a type of universality, some critical exponents do depend on the distribution of disorder.
\end{abstract}
\maketitle


Glassy systems, characterized by extremely slow relaxation and
resultant complex hysteresis and memory effects, are difficult to
study because their dynamics encompass a great range of time scales
\cite{REVIEW}. Glassy materials include those without intrinsic
disorder, such as silica glass, and those where quenched disorder
influences the active degrees of freedom. An example of a model
of the latter is the Edwards-Anderson spin glass model \cite{EA},
which includes the disorder and frustration necessary to capture
many of the complex behaviors seen in disordered magnetic materials.
Though this prototypical model of glassy behavior
was originally proposed well over 30 years ago, many aspects of it remain
poorly understood.  The droplet and replica-symmetry-breaking pictures
of spin glasses provide distinct views of spin glass behavior \cite{YoungBook}.
Analytical results are rare, so numerical approaches are invaluable for both
testing and motivating new ideas.
But the numerics are also exceedingly difficult:
computing spin glass ground states is in general a NP-hard problem \cite{Barahona}.
It is believed that these classes of problems require
exponential computational time to solve exactly \cite{computerScience}.

A fortunate special case which is not prone to this computational
intractability is the two-dimensional Ising spin glass (2DISG).  The
Hamiltonian is $\mathcal{H} = \sum_{\langle ij \rangle}
J_{ij} s_{i} s_{j}$, where the couplings ${\mathcal J}=\{J_{ij}\}$ are independent
random variables coupling classical spins $s=\pm 1$ at
sites $i$,$j$ on a 
square toroidal grid with $L^2$ sites.
The randomness of the sign of $J_{ij}$ leads to competing interactions,
not all of which can be satisfied. In general, such models have
complex (free) energy landscapes and very slow dynamics.  The most
commonly-used distributions for the $J_{ij}$ are the $\pm J$
distribution where each bond value is $\pm 1$ with equal
probability, or the Gaussian distribution where the $J_{ij}$ are
chosen from a univariate Gaussian distribution with zero mean.
One apparent difficulty in using the 2DISG as a model system is that
truly long-range spin-glass order only occurs at zero temperature.
However, at low enough temperature 
such that the correlation length $\xi$  
exceeds $L$ one can study a regime of glassy behavior.
In this glassy regime, the dominant spin configurations are very
sensitive at large length scales to small changes in temperature or other global
perturbations: this sensitivity is referred to as ``chaos''.

Highly developed numerical algorithms
\cite{Barahona,GSKastCities,SaulKardar,GLV,ThomasMiddletonSampling}
can efficiently circumvent the complexity due to disorder and frustration:
ground states and finite-temperature partition functions of the 2DISG may be
computed in time polynomial in $L$.
Still, this model is difficult to study numerically due to strong
finite-size effects. Prior to the present work,
the combination of scaling ideas and exact and Monte Carlo numerical
evidence had been unable to develop a
full understanding of the low-temperature glassy regimes \cite{KatzgraberLeeCampbell}.

In this Letter, we deduce results for the thermodynamics and droplet scaling in the glassy regime
and the accompanying chaos by carrying out precise calculations at low
temperatures and analyzing a variety of quantities, including the sensitivity
of the entropy and energy to boundary conditions.  These results give a
much more consistent and detailed picture of this important model system and
also have applications to other models for disorder in all dimensions. We extend fast
algorithms \cite{SaulKardar, GLV} to study general disorder distributions. By employing arbitrary
precision arithmetic, we have derived reliable numerical results to very low
temperatures and large $L$ (beyond those attained with Monte Carlo simulations).  The ground
states for $\pm J$ vs.\ Gaussian couplings are known to have very different properties
\cite{AMMP,CHK}, but it has been shown \cite{Jorg-etal} that there is in a certain sense
universal behavior, independent of disorder distribution, at finite
temperatures.  The critical behavior
is universal at large length scales where the effective couplings are
continuously distributed \cite{Jorg-etal}, but we argue that 
this scaling and universal chaos 
applies only above a temperature-dependent crossover length
scale $\ell_x(T)$ that itself scales differently for $\pm J$ vs.\ continuously-distributed bare couplings.
We thereby deduce related but different critical exponents for Gaussian
and $\pm J$ disorder; Ref.\ \onlinecite{Jorg-etal} had instead suggested the exponents are the same.
We use numerical results to support these conclusions for the 2DISG.  These
insights into chaos and thermodynamics will help the exploration of glassy
dynamics \cite{Patch} and extend the utility of this model as a standard for
studying spin glasses in general.

Ordering in the glassy regime is subtle, as spin correlations have sample- and location-dependent signs.
However, the magnitude of $\langle s_i s_j\rangle^2$ decays slowly for $|i-j|<\xi$.
This ordering is evident in the effect of boundary conditions on thermodynamic quantities.
We start by computing the partition
function $Z_{\mathrm P}({\mathcal J})$ in a sample with periodic boundary
conditions. Without additional computational effort, we then also have
$Z_{\mathrm{AP}}$ for antiperiodic boundary conditions, where the horizontal
bonds along a vertical column have $J_{ij}$ negated, as both are
signed sums of four Pfaffians
\cite{ThomasMiddletonSampling}. The sample-dependent difference in free energy
$F$ is $\delta F({\mathcal J})=-kT\ln[Z_{\mathrm P}({\mathcal
J})/Z_{\mathrm AP}({\mathcal J})]$.  We use finite differences over $T$
to compute the heat capacity $C$, the average energy
$E$, and the entropy $S$. The sample variances $\mathrm{Var}(\delta X)$ of
$\delta X\equiv X_{\mathrm{P}}-X_{\mathrm{AP}}$ for $X=F$, $E$, $S$, or $C$
help characterize the $\delta X$ distributions.
In the glassy regime $L<\xi$, these differences arise from scale-$L$ relative
domain walls that cross the sample: the magnitudes of $\delta X$ for these walls
are taken to scale as would general droplet
excitations at scale $L$ in an infinite-size system \cite{FisherHuse}.
We study from about $500$ to $10^4$ samples at each
temperature.  The computation of $Z$ for a $512^2$ sample at
$T=0.1$ requires $1.6$ GB of memory  and $4.4\,\mathrm{h}$ on a
2.6 GHz Opteron core. Error bars in all plots in this paper represent
$\pm 1\sigma$ statistical errors.

{\bf Temperature chaos} One of the most notable features of
such a disordered magnet is ``chaos'': a great sensitivity of
the dominant spin configurations to small perturbations of the
temperature \cite{BrayMoore, FisherHuse}.  This sensitivity results
from the delicate balance between entropy and energy in the glassy
state: small temperature shifts 
can flip the sign of the free energy of large-scale droplets.
We define a crossover length $\ell_x(T)$ to be the scale
where the entropy of a droplet or domain wall becomes important: For scales $\ell$ with $\xi(T)>\ell>\ell_x(T)$,
the entropy change $|\delta S|$ is typically larger than $|\delta F|/T$.
A temperature change of size $\delta T < T$ can then reverse
the sign of the free energy $\delta F = \delta E - T \delta S$
of the droplet or domain wall, so this is the chaotic glassy regime.
Here $\delta S$ is continuously distributed, so even if the distribution
of $\delta E$ is discrete, as in the $\pm J$ model, the effective
couplings $\delta F(\ell)$ are continuously distributed for
scales $\ell > \ell_x$.
The natural assumption we make and verify is that when $\ell > \ell_x$ the
exponents for the scaling with $\ell$ are independent of the bare disorder distribution.
In a region of size $\ell$, for $\xi(T)>\ell>\ell_x(T)$ the 
lowest free energy droplet of size of order $\ell$ has a typical free energy
$\delta F(\ell) \sim (\ell/\ell_x)^{\theta}\delta F(\ell_x)$, where
$\theta < 0$ is the usual stiffness exponent.
The boundary of this droplet is a fractal domain wall with dimension $d_f$, and
the typical $\delta S$ scales as
$\delta S (\ell) \sim ( \ell/\ell_x)^{d_f/2} \delta S(\ell_x)$.
In verifying this picture, we will use accepted values $\theta\cong -0.28$ and
$d_f\cong1.274$ \cite{AMMP,CHK,MelchertHartmann} computed in ground state simulations
with Gaussian couplings, rather than
fitting our data to redetermine these values.

\begin{figure}[tb]
\centering
\includegraphics[width=3.4in]{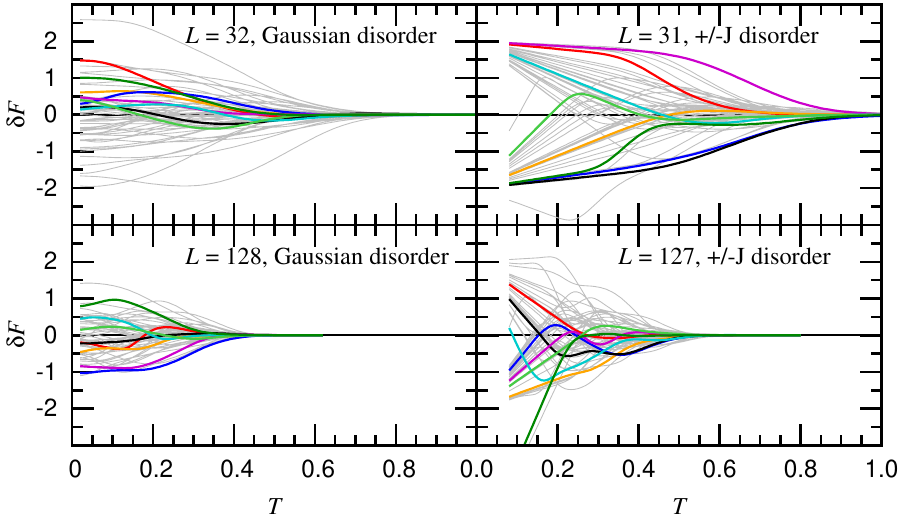}
\caption{ [color online] Chaos is evident in plots of $\delta F\equiv F_{\mathrm P}-F_{\mathrm{AP}}$, the change in free energy
   with boundary conditions,  vs.\ temperature $T$ for both Gaussian and $\pm J$ disorder.
   Each panel shows curves for 50 samples of size $L^2$.
   Randomly chosen curves are highlighted to improve visibility. The number of zero crossings of $\delta F$ increases with $L$, even
   over the diminishing temperature range where $\delta F$ is appreciable.}
\label{fig:deltaF}
\end{figure}

Chaos can be seen in zero crossings of $\delta F(T)$,
which imply a reordering of the values of
$F$ for P and AP boundaries at scale $L$.  A subset of the data for
$\delta F (T)$ is displayed in Fig.\ \ref{fig:deltaF}: each curve indicates
$\delta F(T)$ for a single sample (also see \cite{Sasaki-etal}).  Note that in
the region of appreciable $\delta F$, the number of crossings increases as $L$
increases.  This helps justify the study of the 2DISG as a chaotic glassy model
since a large sample will go through a large number of
very different states as the temperature is lowered.

{\bf Non-universal scaling and the crossover length}
We argue that the dependence of the crossover length $\ell_x$
on temperature $T$ is non-universal, so that some of the 
critical exponents for different disorders are distinct, though related.
At short length scales, $\ell<\ell_x(T)$, $T\delta S < \delta F$, so that the
thermodynamics at these scales is determined primarily by energetics and the
$T=0$ fixed point sets the scaling behavior. In a system of size $L$ where $\ell_x(T) > L$,
chaos is frozen out (there are typically no zero crossings in $\delta F$).
This small-$\ell$ or low-$T$ nonchaotic regime is what is seen in $T=0$
simulations.  It is characterized by marked differences in behavior between the
$\pm J$ and Gaussian disorder distributions \cite{AMMP,CHK}.

For Gaussian disorder, the entropy of a droplet at scale $\ell < \xi$ is 
concentrated 
on the fractal domain wall with a typical total length $\ell^{d_f}$.
The entropy difference is due to the difference in the local excitations that
are affected by the introduction of this domain wall.  Presuming that the local
excitations at length scale $\sim 1$ have a gapless spectrum with a constant width,
the fraction of the domain wall that is thermally
active and contributes to the entropy difference is proportional to
$T$.  The entropy of the droplet, $\delta S$ is then a sum of $\sim T\ell^{d_f}$ terms of
random sign so that $\mathrm{Var}(\delta S) \approx T \ell^{d_f}
\sigma(T\ell^{-\theta})$ with the scaling function 
$\sigma(x)\rightarrow \mathrm{const.}$ as $x\rightarrow 0$.  The non-chaotic
regime breaks down when $T\delta S\sim T^{3/2} \ell_x^{d_f/2} \sim \ell_x^\theta \sim \delta F$, giving
$\ell_x \sim T^{-3/(d_f - 2 \theta)}$ for Gaussian disorder.

For $\pm J$ disorder and low $T$, the entropy of a scale-$\ell$ droplet with $\ell < \ell_x(T)$ is found to scale as
$\delta S \sim \ell^{\theta_S}$. The value of $\theta_S$ is fixed by the $T=0$
domain wall entropy due to zero-energy spin rearrangements, and
estimated numerically to be $\theta_S \cong 0.5$ \cite{LukicEtAlJStat}.
Additionally, the typical (free) energy of excitations at zero temperature has
been found to be $\mathcal{O}(1)$, independent of $L$ \cite{AMMP,CHK}, which we have
verified: fitting $\delta F(T=0)$ to
a simple power law over $L=32\rightarrow 256$ gives an exponent with magnitude less than $0.015$.
The crossover scale $\ell_x$ therefore occurs when $\delta
E\sim \delta F \sim T\delta S \sim \mathcal{O}(1)$, giving
$\ell_x \sim T^{-1/\theta_S}$, so that Gaussian and $\pm J$ distributions have
distinct scaling.  Crucially, this also means $\delta F(\ell_x) \sim
\mathcal{O}(1)$, in contrast with $\delta F(\ell_x) \sim \ell_x^{\theta}$ for the
Gaussian case; this introduces an additional source of nonuniversality. 

For Gaussian disorder, the free energy magnitude scales as
$\mathrm{Var}(\delta F) \approx \ell^{2\theta} \phi(T \ell^{-\theta})$ with scaling function
$\phi(x)\rightarrow \mathrm{const.}$ as $x\rightarrow 0$
\cite{BrayMoore,FisherHuse}.  The correlation length $\xi(T)$ is set by
the scaling argument $x=\mathcal{O}(1)$, which gives $\xi\sim T^{-\nu_G}$ with
$\nu_G   =-1/\theta\cong 3.5$.  For $\pm J$
disorder, low $T$ and $\ell>\ell_x(T)$ in the universal, chaotic regime,
$\mathrm{Var}(\delta F) \approx (\ell/\ell_x)^{2\theta} \phi[T ( \ell / \ell_x )^{-\theta}]$
with the same universal scaling function $\phi(x)$.  But here the argument of the scaling function has
different $T$-dependence from what it has with continuously-distributed disorder.
As a result the correlation length scales with a different exponent, $\xi\sim T^{-\nu_\pm}$,
with $\nu_\pm=1/\theta_S - 1/\theta\cong 5.5$.

\begin{figure}[h]
\centering
\includegraphics[width=3.4in]{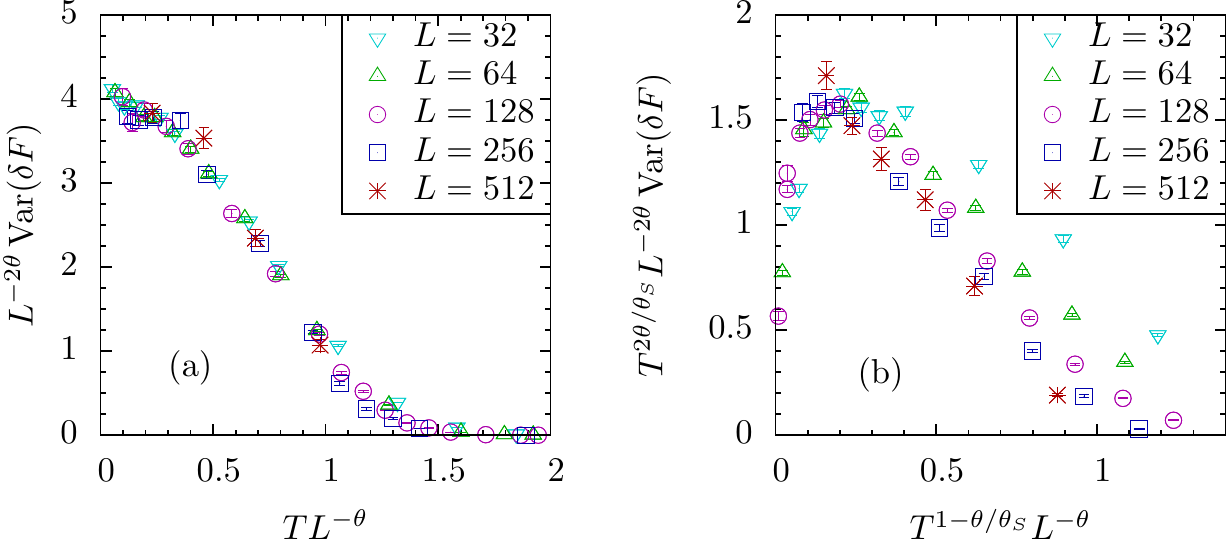}
\caption{ [color online] (a) Scaling collapse for the variance of the twist free energy,
  $\mathrm{Var}(\delta F)$, for Gaussian disorder, with
  $\mathrm{Var}(\delta F)L^{-2\theta}$ plotted vs.\ $TL^{-\theta}$.
  (b) Scaling collapse for $\pm J$ disorder that accounts for the crossover-length $\ell_x(T)$
  so that convergence to the same universal scaling function is expected (up to
  overall scales for $T$ and $L$). There are no fitted exponents in these plots: accepted
  values $\theta=-0.28$ and $\theta_S=0.50$ are used.}
\label{fig:F}
\end{figure}

We computed $\mathrm{Var}(\delta F)$ for samples chosen independently at each
temperature. We directly test the scaling forms expected for $\delta F$ by a
finite-size scaling collapse of this data with no free parameters.
For the Gaussian data, shown in Fig.~\ref{fig:F}(a),
we find agreement with the expected exponents and thus a good estimate
of the scaling function $\phi(x)$.  We plot the data for the $\pm J$ model
using its expected scaling in Fig.~\ref{fig:F}(b), where the same scaling
function $\phi(x)$ with only constant rescaling should appear.
Although the expected deviations are apparent at low temperature and small $L$ where $L<\ell_x(T)$
and at $T=\mathcal{O}(1)$, we see a trend towards good collapse of the data for larger $L$, consistent with
the proposed new scaling, and similar curves for the $\phi(x)$.
It appears that for $\pm J$ disorder the
universal regime of $L>\ell_x(T)$ and $T<\mathcal{O}(1)$ does not emerge until
$L\apprge 100$.

\begin{figure}[h]
\centering
\includegraphics[width=3.2in]{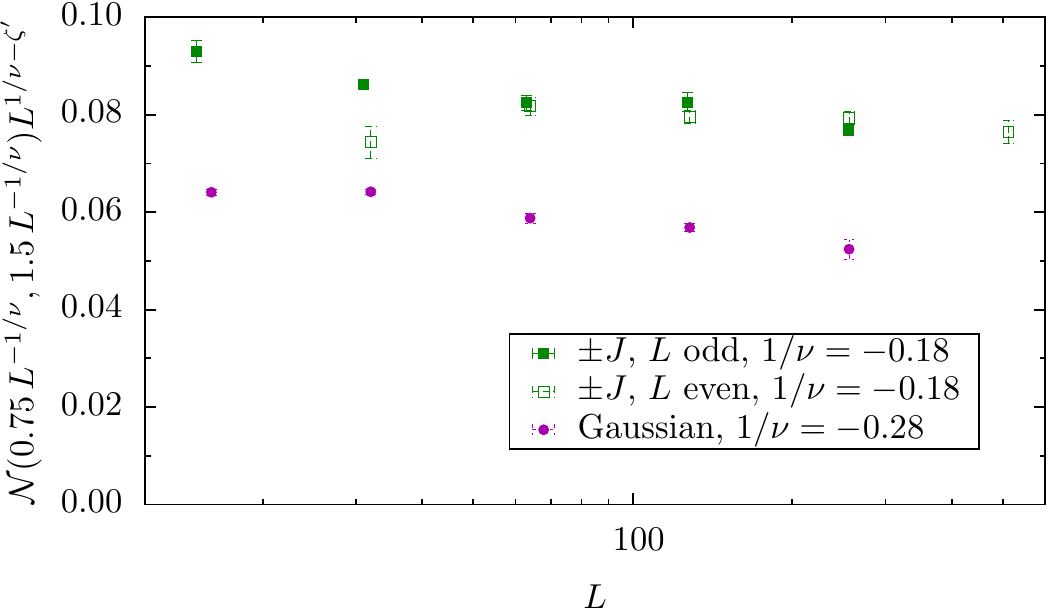}
\caption{[color online] A plot of the scaled crossing rate
$L^{-\zeta'}\mathcal{N}(T_1,T_2)/(T_2-T_1)$, where the average number  of
zero crossings for $\delta F$ in the interval $T_1< T<T_2$ is $\mathcal{N}(T_1,T_2)$ \cite{HuseKo},
using $(T_1,T_2)=(0.75\,L^{-1/\nu},1.5\,L^{-1/\nu})$. The rate is expected to scale
with $L$ as $\propto L^{\zeta'}$, with $\zeta'_G=0.78(1)$ and $\zeta'_\pm=0.77(2)$.
The expected nearly constant scaled rate is seen for both Gaussian and bimodal disorder.}
\label{fig:chaos}
\end{figure}

{\bf The chaos exponent}
The rate of change of $\delta F$ with $T$ in an individual sample is given by the entropy difference, 
$\delta S=-\frac{\partial\, \delta F}{\partial T}$.  Thus, in the chaotic regime, the temperature change needed to
change the sign of $\delta F$ is typically $\Delta T\sim\delta F/\delta S$.  Using the above expressions for
$\delta F$ and $\delta S$, the rate of sign changes in $\delta F$ as $T$ is varied for Gaussian disorder is given by
$|\delta S|/|\delta F|\sim L^{\zeta'_G} \kappa_G(TL^{1/\nu_G})$ with chaos exponent
$\zeta'_G=(d_f+1/\nu_G)/2$ and scaling function $\kappa(x) \propto x^{1/2}$ at small argument.  For
$\pm J$ disorder, $T\delta S(\ell_x) \sim \mathcal{O}(1)$, so that in
the chaotic regime $\delta S \sim
(\ell/\ell_x)^{d_f / 2}/T$ and the rate of sign changes is $|\delta S|/|\delta F|\sim
(L/\ell_x)^{d_f/2-\theta}T^{-1}\kappa_\pm(TL^{1/\nu_\pm})$, with $\kappa_\pm$
constant for small arguments.  This gives $\zeta'_\pm =
(d_f/2-2\theta)[\theta_S/(\theta_S-\theta)]$.  Though these two expressions for
$\zeta'$ are quite different, they are predicted to have similar values:
$\zeta'_G=0.78(1)$ and $\zeta'_\pm=0.77(2)$ (Fig.\ \ref{fig:chaos}).


{\bf Specific heat} 
The behavior of the specific heat $C$ for $T\rightarrow 0$ in a 2DISG is
dominated by the smallest thermally active droplets that have nonzero
energy \cite{FisherHuse}.  If the disorder distribution is continuous,
these are the smallest droplets of size $\mathcal{O}(1)$
and energy of order $T$.  They have a density proportional to $T$ and
contribute a linear term in the low $T$ specific heat: $C \sim T$.

\begin{figure}[h]
\centering
\includegraphics[width=3.3in]{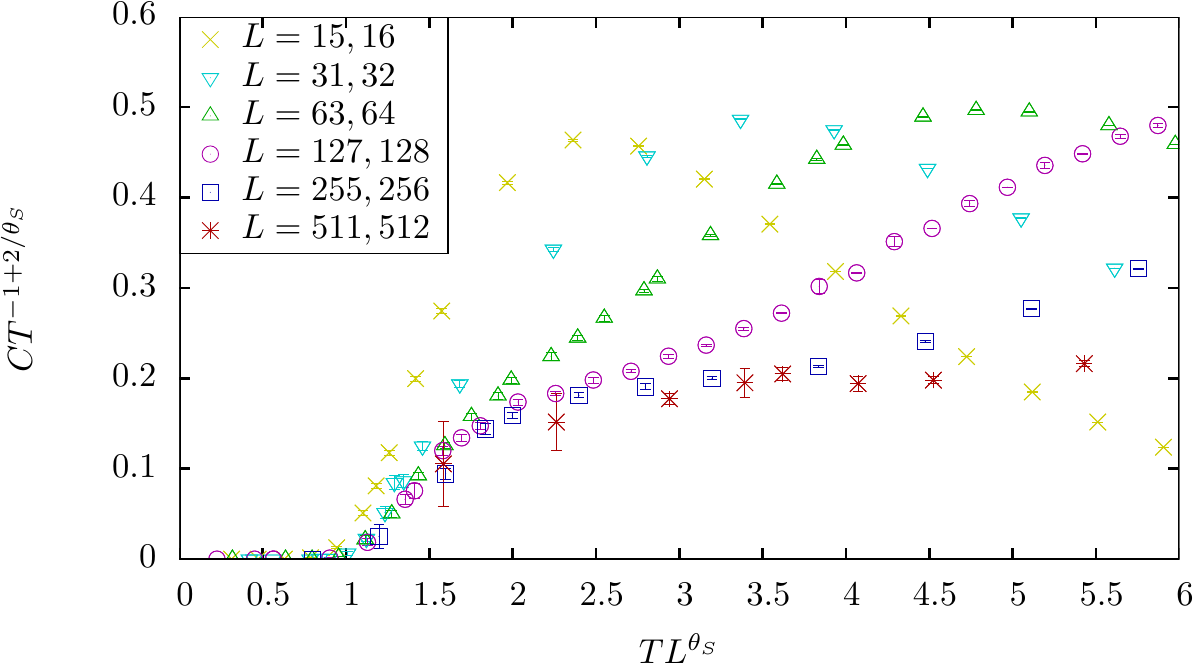}
\caption{ [color online] A scaling plot for the heat capacity $C$ for the $\pm J$ spin glass. The scaled
heat capacity per spin, $C T^{1-2/\theta_S}$ is plotted as a function of the scaled temperature
$T L^{\theta_S}$ for samples of size $L$ with the value of $\theta_S = 0.50$ consistent
with the zero temperature scaling of the entropy of domain walls, $\delta S \sim L^{\theta_S}$ \cite{LukicEtAlJStat}.
This collapse is consistent with the ``bulk'' heat capacity being due to excitations of minimal
size $\ell_x\sim T^{-1/\theta_S}$. }
\label{fig:specheatPMJ}
\end{figure}

For $\pm J$ disorder, on the other hand, the droplets with the lowest nonzero energy have
$\delta E = \mathcal{O}(1)$.  Such droplets with size $\ell<\ell_x(T)$ have $\delta F > T$ at low $T$
and are not thermally active.  Thus at low temperature $T$ the smallest active droplets with nonzero energy
are of size $\ell_x(T)$.  These active droplets each contribute $\sim 1/T^2$ to the specific heat
and have density $T/\ell_x^2(T)$ so the specific heat scales as $C\sim T^{2/\theta_S-1}$.
In an a average over finite-sized samples this power-law specific heat is cut off
at the lowest temperatures when the size $\ell_x(T)$ of
these active droplets exceeds $L$.  This produces the low-$T$ finite-size scaling
form $C\approx T^{2/\theta_S-1}c(TL^{\theta_S})$,
where $c(x)$ is a scaling function that goes to a constant for large
argument (for temperatures where $\ell_x(T) \ll L \ll \xi(T)$).  Our $\pm J$ specific heat data are
shown in this scaling form in Fig. \ref{fig:specheatPMJ}.  The intermediate temperature regime that
corresponds to the power-law specific heat only appears for $L>100$.
In fact, our data for intermediate temperatures, $T \approx 0.35$,
are entirely consistent with previously published work \cite{KatzgraberLeeCampbell}, which saw an
effective exponent $\alpha\approx -4.2$, with $C\sim T^{-\alpha}$, but the effective exponent crosses
over to lower values as $T$ decreases.



In addition to the leading terms that we discuss above and detect in our 
numerical results,
there are weaker singular contributions to $C$ that
result from the diverging $\xi(T)$ as $T\rightarrow 0$
that are also there in principle, although they will be 
extremely difficult to detect.
Standard hyperscaling at $T=0$ predicts a contribution that scales as
$C\sim T^{-\alpha} \sim T^{2\nu}$ \cite{Jorg-etal}, but the chaos
changes this, making this contribution larger at low $T$.
The contribution to the specific heat from droplets of scale $\ell\le\xi(T)$ is
the product of their density $\sim 1/\ell^2$, the fraction that are thermally
active $\sim T/\delta F(\ell)$ and the average contribution of such an active droplet
to the heat capacity $\sim (\delta E_\mathrm{act}(\ell)/T)^2$.  For standard hyperscaling, the last two
factors are of order one for $\ell=\xi(T)$, but the chaos instead makes the last term larger.
The subdominant contribution to $C$ for Gaussian disorder from scale $\xi$ is
$C_\mathrm{sing}\sim T^{2\nu_G(1-\zeta'-\theta)}$, since $\delta E_{\mathrm act}/T \sim \delta S_\mathrm{act} \sim \ell^{\zeta'+\theta}$.
However, the first factor always wins, keeping the smallest thermally active droplets with
$\delta E > 0$ dominant in the specific heat.


{\bf Discussion} Precise numerical calculations have allowed us to test new
scaling relations for the
thermodynamics of the 2DISG in detail and to clearly demonstrate non-universality and
study its origin.
The large values of $\nu$ (and long crossover lengths for $\pm J$ disorder)
necessitate using large systems to see the scaling behavior.
Even though we assumed that scaling in the chaotic regime is weakly universal
with the same values of $\theta$ and $d_f$ at any given $T$,
our results show that the critical exponents are nevertheless
different for Gaussian and $\pm J$ disorder, due to temperature- and disorder- dependent
crossover length scales.
We predict a violation of hyperscaling due to chaos, which is a general
phenomenon that is present in higher dimensions as well.
These results will lead to further work on the thermodynamics
and glassy dynamics of this glassy model at low temperature.

This work was supported in part by NSF grants DMR-1006731 and DMR-0819860. We are
grateful for the use of otherwise idle time on the Syracuse University
Gravitation and Relativity computing cluster, supported in part by NSF
grant PHY-0600953, to generate almost all of the
numerical results, and additional time on the Hypatia cluster (supported in part by NSF
DMR-0645373) and the Brutus cluster of ETH Zurich.  We thank Helmut Katzgraber for helpful comments.


\end{document}